\documentclass[aps,twocolumn,showpacs,preprintnumbers,amsmath,amssymb,prx]{revtex4-1}
\usepackage{epsf,amsmath,amssymb,verbatim,color,multirow,pifont}
\usepackage{graphicx,times,mathptm,amsmath}
\usepackage{braket,commath,epstopdf}
\usepackage[usenames,dvipsnames,svgnames,table]{xcolor}
\usepackage{soul}

\begin{document}

\title{Dismantling efficiency and network fractality}
\author{Yoon Seok Im}
\author{B. Kahng}
\email{bkahng@snu.ac.kr}
\affiliation{$^1$CCSS, CTP and Department of Physics and Astronomy, Seoul National University, Seoul 08826, Korea}

\date{\today}

\begin{abstract}
Network dismantling is to identify a minimal set of nodes whose removal breaks the network into small components of subextensive size. Because finding the optimal set of nodes is an NP-hard problem, several heuristic algorithms have been developed as alternative methods, for instance, the so-called belief propagation-based decimation (BPD) algorithm and the collective influence (CI) algorithm. Here, we test the performance of each of these algorithms and analyze them in the perspective of the fractality of the network. Networks are classified into two types: fractal and non-fractal networks. Real-world examples include the World Wide Web and Internet at the autonomous system level, respectively. They have different ratios of long-range shortcuts to short-range ones. We find that the BPD algorithm works more efficiently for fractal networks than for non-fractal networks, whereas the opposite is true of the CI algorithm. Furthermore, we construct diverse fractal and non-fractal model networks by controlling parameters such as the degree exponent, shortcut number, and system size, and investigate how the performance of the two algorithms depends on structural features.
\end{abstract}

\maketitle

\section{Introduction}
Network science concerns phenomena in systems of multiple nodes interacting with each other through links. In heterogeneous networks, whose nodes have various number of connections and hierarchy, identification of influential nodes is an important issue. If we know the characteristics of nodes that play crucial roles in a specific dynamic process, we can control the process by modifying the connectivity of the network. For instance, one may wonder how to identify super transmitters, who are likely to induce an epidemic outbreak when they are infected. Once the super transmitters are identified, the spread of disease can be suppressed by vaccinating or quarantining them first. Because disease epidemics can spread rapidly on networks and be fatal to humans, many studies have been performed to identify super transmitters using epidemic models such as the susceptible-infected-recovered (SIR) model \cite{Liu2016, Chen2013, Chen2012, Yeruva2016, Deng2016, nphys2010}. As in the prevention of epidemics, modification of a network requires resources and time. Thus,  optimization of influence parameters by selecting an appropriate set of nodes is essential when propagation processes must be controlled on complex networks. It can be used in any attempt to control the behavior of an entire network using limited resources, such as viral marketing \cite{Richardson2002}, political campaigns, and military intelligence \cite{Lokhov2007}.

In optimal percolation, also called network dismantling, a minimal set of nodes is identified, whose removal breaks the giant connected component of a network into small components of subextensive size. It can be mapped to optimal immunization and the spreading problem \cite{Makse2015}. Practically, optimal percolation offers a general countermeasure against infectious disease, no matter how contagious it is, where the size of the giant component is an upper bound on epidemic outbreak \cite{Grassberger2016}.
Optimal percolation is an NP-complete problem, like the other optimal influence problems \cite{Kempe2003}. One cannot expect a deterministic algorithm to work within polynomial-time complexity unless the answer to the famous P-NP problem is proved to be affirmative \cite{Karp1972}. Instead, the solution can be chosen from among the candidates by guessing and then checked in polynomial time. Given a finite fraction as the number of nodes to be removed, we can find the exact optimal percolation set with the given size by checking every possible candidate. However, as the system size increases, the number of cases increases exponentially. Thus, to deal with large networks, we need to develop a method to guess a good candidate for the solution at a scalable time complexity. 

As the first step, one can delete important nodes in turn, using centrality measures such as the degree, eigenvector centrality, or closeness centrality as a criterion for the importance \cite{Lu2016}. For many graph instances, the performance can be enhanced by recalculating the centrality measures after each removal \cite{Holme2002}. However, a centrality measure does not guarantee the importance of a collection of nodes as a set, even if each chosen node is important by itself. Even if we choose a good centrality measure while trading off appropriately between the scalability and accuracy, its effectiveness depends heavily on the topology of the target network. Traditional optimization methods such as the Monte Carlo method take a very long time to approach the optimal value, and greedy algorithms give unreliable results in many cases \cite{Altarelli2014}.

As alternatives, several heuristic algorithms, theoretically based on belief propagation (BP) \cite{Yedidia2004}, have been proposed recently, including those designed to find the dismantling set, that is, the minimal set of nodes that can be removed to break the giant connected component into small pieces of subextensive size. In BP, global information is transmitted by iteration of message-passing equations for local quantities. This characteristic is appropriate for the optimal percolation problem, where we should consider the global influence of node removal, although we have to keep the quantities local because we need a scalable algorithm. In fact, most state-of-the-art algorithms use a BP method based on the spin glass theory in statistical physics and some characteristics described by graph theory. The BP-based decimation (BPD) algorithm \cite{Zhou2016} shows outstanding performance among heuristic algorithms for most graph instances. The min-sum algorithm \cite{Braunstein2016}, the performance of which is known to be comparable to that of the BPD algorithm, is also based on BP. The collective influence (CI) algorithm \cite{Makse2015}, one of the algorithms based on centrality, also starts theoretically by considering the stability of message-passing equations, although it is approximated to use a centrality measure so-called the CI. The three algorithms mentioned above are very scalable and are known to work in $\mathcal{O}(N \log N)$ time complexity.

The optimal percolation problem is deeply related to the characteristics of loops. Many dismantling algorithms assume that the network is locally treelike, that is, the number of local loops is negligibly small. Under this assumption, it can be shown that the minimal dismantling set coincides with the minimal decycling set, that is, the minimal set of nodes whose deletion leads to the removal of every loop in the graph~\cite{Braunstein2016}. Those algorithms in Refs.~\cite{Zhou2016, Braunstein2016} use this fact as a first step in dismantling: First, remove loops from the network, and then break down the remaining trees into small pieces. Furthermore, the validity of the BP method relies on the loop characteristics of the network. The BP method is exact on tree graphs, and it gives a good approximation if the correlation between neighbors of a node is sufficiently small in the cavity graph~\cite{Mezard}. This condition is realized if there is no local loop, i.e., the network is locally treelike. However, algorithms based on the BP method are reportedly still effective on real-world networks that contain many local loops~\cite{Zhou2016, Braunstein2016, Zdeborova2014}.

The loop characteristics of networks have been categorized quantitatively by the fractal scaling property~\cite{Goh2006}. Fractal scaling represents the power-law relation between the minimum number of boxes $N_B$ to cover the entire network and the size of the boxes $\ell_B$, $N_B(\ell_B)\sim \ell_B^{-d_B}$ with a finite fractal dimension $d_B$~\cite{song}. It has been observed, however, that not all networks are fractals, and most of the random network models proposed to date are also not fractals. Here we aim to characterize the effectiveness of dismantling algorithms on loopy graphs in the perspective of the fractality of networks. We find that the BPD algorithm works more efficiently for fractal networks than for non-fractal networks, whereas the opposite is true of the CI algorithm. Moreover, the performance gap between the two algorithms is smaller for non-fractal networks than for fractal networks.

This paper is organized as follows: We show the dismantling performances of the BPD and CI algorithms on the two real-world newtorks, the World Wide Web and Internet, in Sec. II. Similar works are performed on the model networks with various structural features in Sec. III. How the performance relies on the structural features is also discussed. In Sec. IV, we reproduce the dismantling performances of the real-world networks by controlling structures of model networks, and discuss its implication. The final section is devoted to summary and discussion.

\begin{figure}[t]
\includegraphics[width=\linewidth]{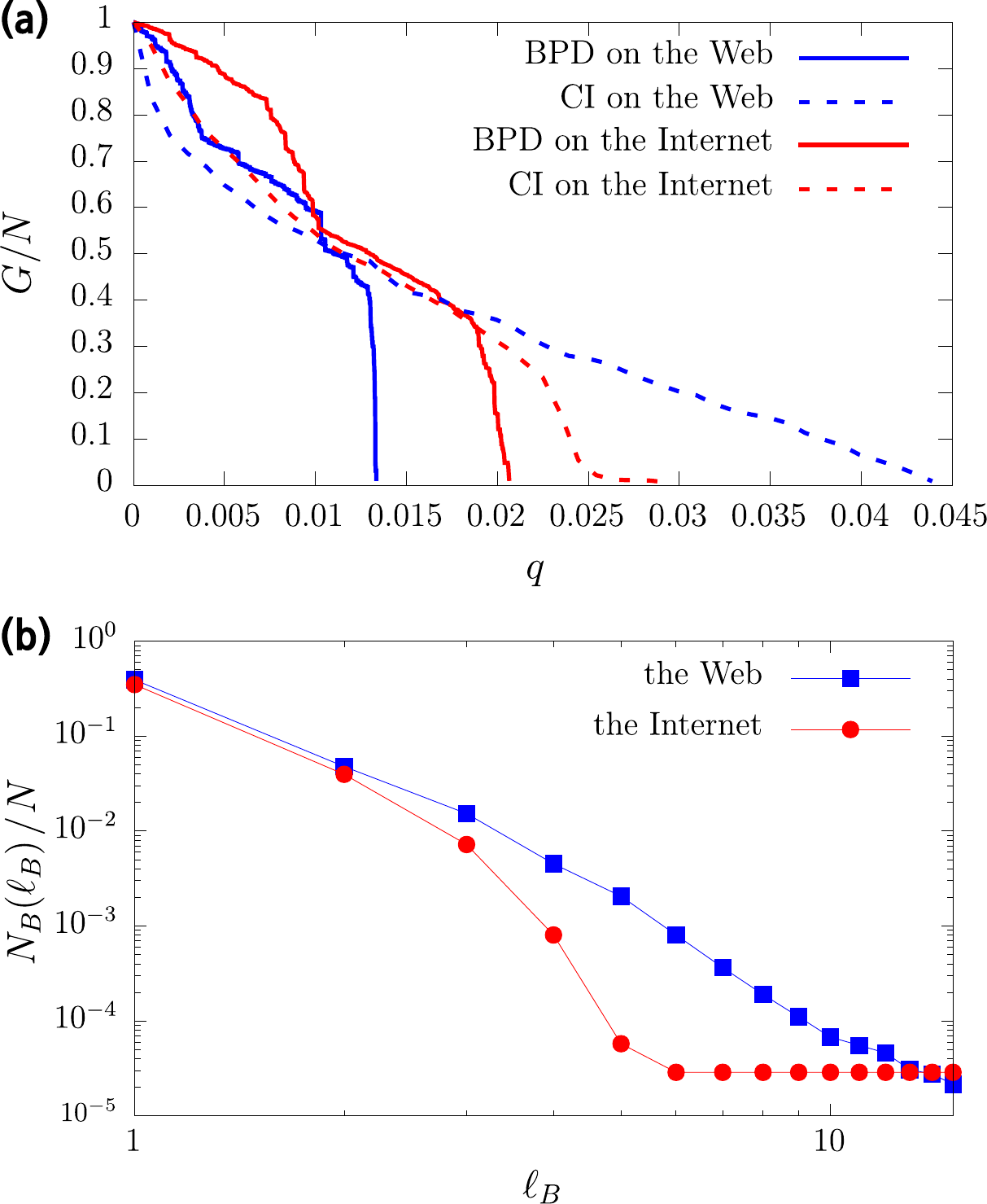}
\caption{(a) Performances of the BPD and CI algorithms for two real-world networks. As the fraction $q$ of deleted nodes increases, the fraction of giant connected component $G$ to the system size $N$ decreases. (b) Fractal scalings of the two real-world networks, the World Wide Web and Internet at the autonomous level, measured by a random sequential box-covering method \cite{Kim2007}. The web appears to be a fractal, whereas the Internet is not.}
\label{fig:comp-real}
\end{figure}

\section{Performances on real-world fractal and non-fractal networks}

Networks can be factored into a skeleton and shortcuts. The skeleton is a spanning tree formed by $N-1$ edges with the highest betweenness centrality~\cite{betweenness1,betweenness2} or load~\cite{load}, and shortcuts connect different branches of the tree, forming loops of various sizes. In particular, a skeleton formed by the critical branching process, in which the mean number of offspring is unity, seems to be required for a network to be fractal~\cite{Goh2006,burda}. If the shortcuts are local, the fractality of the skeleton is preserved, and the resulting network is still fractal. If we add global shortcuts, they deform the fractal scaling behavior of the skeleton, and the network is non-fractal. As the performance of BP-based algorithms relies on the number of local and global loops, it is suggested that the fractality affects the performance of dismantling algorithms.

Figure~\ref{fig:comp-real} supports this suggestion, showing that the order of the resilience of sample networks can be changed if one uses different algorithms to obtain the optimal percolation threshold. One of the sample networks is the World Wide Web~\cite{notre_dame}, which is considered as an undirected network. Two nodes are regarded as connected if there is a hyperlink from one to the other. It is a scale-free network with degree exponent $\gamma \approx 2.6$, and appears to be a fractal~\cite{Goh2006}.
Another sample is the Internet topology at the autonomous level collected in early 2010~\cite{as}. It has a power-law degree distribution with degree exponent $\gamma \approx 2.1$, and it is a non-fractal network \cite{Goh2006}.

Although the BPD algorithm performed better in both networks, the performance gap between the BPD and CI algorithms was much smaller on the Internet, which is a non-fractal (Fig.~\ref{fig:comp-real}). The gap increases sufficiently on the web, which is a fractal, to separate the optimal percolation thresholds found by each algorithm from the thresholds for the Internet. In other words, the Internet is more vulnerable than the web to the CI algorithm, but the web is more fragile under the BPD algorithm. 
As these two networks have many different features such as system size, degree exponent, and fractality, we have to narrow down the possible factors that affect the gap between the performance of the two algorithms. 
We need a method to generate model networks with the desired topological characteristics such as system size, degree distribution, and fractality.

\begin{figure}[t]
\centering
\includegraphics[width=\linewidth]{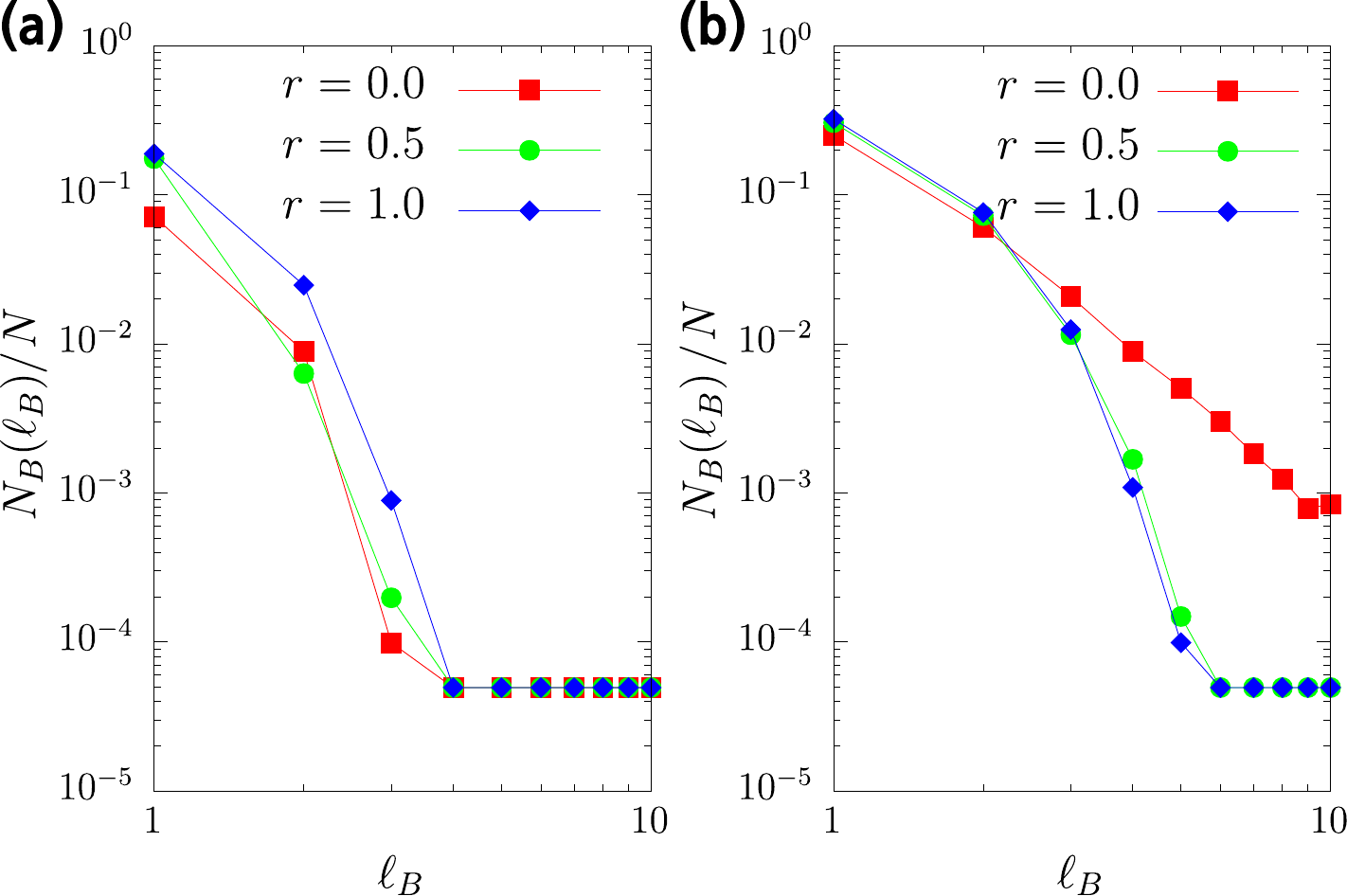} 
\caption{Fractal scaling behaviors of the fractal model networks measured by the so-called random sequential box-covering method \cite{Kim2007}. Networks (a) and (b) are generated on the basis of critical branching trees with degree exponent $\gamma=2.1$ and $\gamma=2.7$, respectively. On these skeletons, $sL$ shortcuts are added within the hopping distance $d=8$, where $s=3.0$, and $L=N-1$ is the size of the critical branching tree. The network size was set to $N=2.0\times 10^4$. Shortcuts were drawn locally at first ($r=0.0$), and a fraction $r$ of the shortcuts were rewired randomly ($r=0.5, 1.0$). For $\gamma=2.1$, the network is non-fractal even when shortcuts are not rewired. For $\gamma=2.7$, the network is fractal when shortcuts are added and becomes non-fractal as the shortcuts are rewired.}
\label{fig:FS-model}
\end{figure}

\section{Performances on fractal and non-fractal model networks} 

The fractal network model (FNM) introduced in Ref.~\cite{Goh2006} can be used to construct model networks with the desired loop characteristics. First, we build a critical branching tree of the desired size with $N$ nodes and $L=N-1$ links. Degree distribution of the critical tree is controlled by the probability $b_m$ of generating $m$ offspring. When $b_m\sim m^{-\gamma}$ with $\sum_m mb_m=1$ is taken, a critical branching tree with the degree exponent $\gamma$ is generated. Its fractal dimension is determined as $d_B=(\gamma-1)/(\gamma-2)$. Then, we add shortcuts to the tree as follows: First, stubs are added to each node, the number of which is proportional to the degree of each node of the critical tree. The total number of stubs is given as $2sL$, where $s$ is a control parameter. Next, we add $sL$ shortcuts between unconnected stubs at different nodes. To maintain the fractal nature, we limit the hopping distance $d$ between nodes that are to be connected by a shortcut. This limitation is required for local loops to conserve the global connectivity and the fractality of the branching tree. A fractal network can be deformed to a non-fractal network by rewiring the fraction $r$ of $sL$ shortcuts without the distance limitation and changing the degree distribution. By rewiring links, shortcuts can be changed to long-range loops, which reduce the network diameter or destroy distinct modules. As more shortcuts are rewired, the network loses more of its modularity, and the fractality is broken further, resulting in a non-fractal network. 

It is noteworthy that when the degree exponent of the branching tree is close to two, the network is more centralized at the hub. Then the network becomes non-fractal even if the shortcuts were added locally and not rewired [Fig.~\ref{fig:FS-model}(a)]. This is because the network diameter is easily reduced by connecting neighbors on different hubs. On the other hand, as shown in Fig.~\ref{fig:FS-model}(b), a critical branching tree with a large degree exponent ($\gamma=2.7$) maintains its fractality when local shortcuts are added and becomes non-fractal when the shortcuts are rewired.


\begin{figure}[t]
\includegraphics[width=1.0\linewidth]{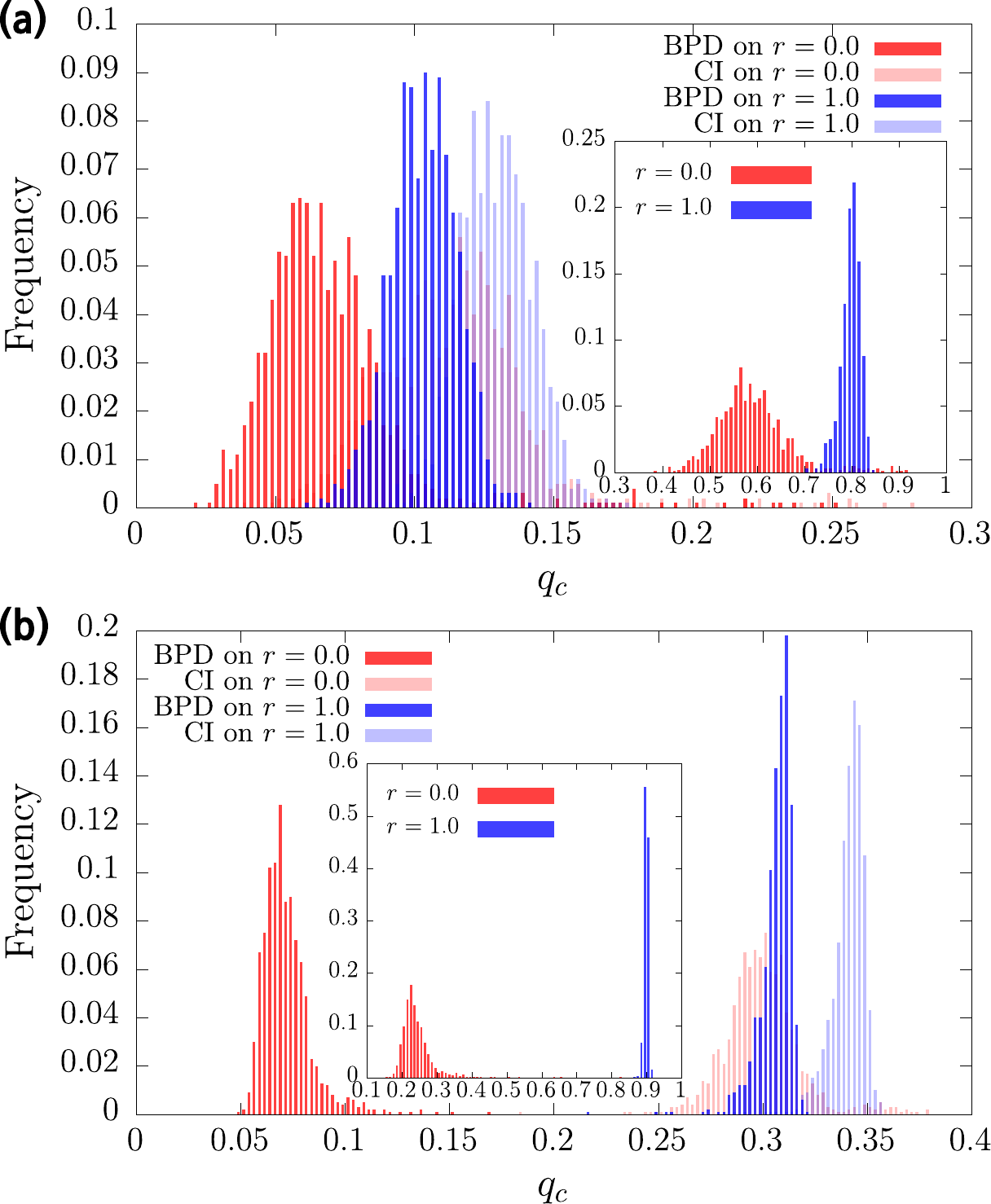}
\caption{Performance comparison of the BPD and CI algorithms on networks with different shortcut rewiring ratios. Each histogram represents a distribution of $q_c$, optimal percolation threshold found by BPD or CI algorithm on the family of networks generated by FNM with specified parameters. (a) non-fractal model networks with $\gamma\approx 2.1$ and (b) fractal model networks with $\gamma\approx 2.7$. Data are obtained from $10^3$ realizations. The number of nodes in each model network ranges between $2\times 10^4$ and $2.2\times 10^4$, and the shortcut parameter is set to $s=3.0$.}
\label{fig:comparison-rewiring-ratio}
\end{figure}

For a given set of parameters such as the degree exponent $\gamma$, the shortcut parameter $s$, and rewiring ratio $r$, we generate $10^3$ individual realizations and obtain the performance of the BPD and CI algorithms for these realizations. Because the algorithms were designed to thoroughly break the giant component into subextensive (small) clusters, they provide good estimations of each optimal value around $G=0$ but can be incorrect far from $G=0$. We present the distributions of the optimal percolation threshold, that is, the proportion of deleted nodes needed to reduce $G$ less than $1\%$ of its original size, rather than showing each curve from $G=1$ to $G=0$, which represents the response of the network to the algorithm.

\subsection{Dependence on shortcut rewiring ratio} 

To find the dependence of the algorithm's performance on the fractality, we fix the parameters used to construct the networks, such as the degree exponent $\gamma$ and shortcut parameter $s$, but control the shortcut rewiring ratio $r$. When $r=0$, we expect fractal networks to be generated. When $r=1$, every added link, i.e., a shortcut, is not limited in distance, and a non-fractal network is generated. However, as we noted previously, the fractality of a model network is broken even at $r=0$ when the degree exponent $\gamma$ is slightly above two. Thus, we compare the performance of the algorithms for networks with $r=0$ and $r=1$ separately on networks where the degree exponent was set to $\gamma\approx 2.1$ and $\gamma \approx 2.7$.  

As hypothesized, the fractality affects the performance of each algorithm. For the model networks with $\gamma\approx 2.1$ (Fig.~\ref{fig:comparison-rewiring-ratio}(a)), the separations between distributions are unclear, and thus they overlap greatly, even though the BPD algorithm is probably better than the CI algorithm for both $r=0$ and $r=1$. 
The superiority of BPD over CI appears in every case observed throughout this study. The insets in the figure show the distribution of the ratio of $q_{c,\text{CI}}$ and $q_{c,\text{BPD}}$, which are measured for each algorithm on the same graph instance. As shown in the insets, there is no single realization in which $q_{c,\text{CI}}$ is less than $q_{c,\text{BPD}}$.
Both algorithms find that fully rewired ($r=1$) networks are harder to destroy than networks with only local shortcuts ($r=0$).
For $\gamma\approx 2.7$ (Fig.~\ref{fig:comparison-rewiring-ratio}(b)), the separation between the distributions of $q_c$ for the BPD and CI algorithms is clear, and those distributions do not overlap greatly. In particular, the performance gap is much larger for the model networks with $r=0$, which are fractals, than for those with $r=1$, which are non-fractals. In the inset, the distribution of the ratio $q_{c,\text{BPD}}/q_{c,\text{CI}}$ differs greatly from the distribution for the rewiring ratio $r=0$ once some rewiring process occurs. This observation also explains the overlap between the BPD and CI algorithms for $\gamma\approx 2.1$, where both $r=0$ and $r=1$ yield non-fractal networks. The BPD algorithm is especially effective on fractal networks.

\begin{figure}[t]
\includegraphics[width=1.0\linewidth]{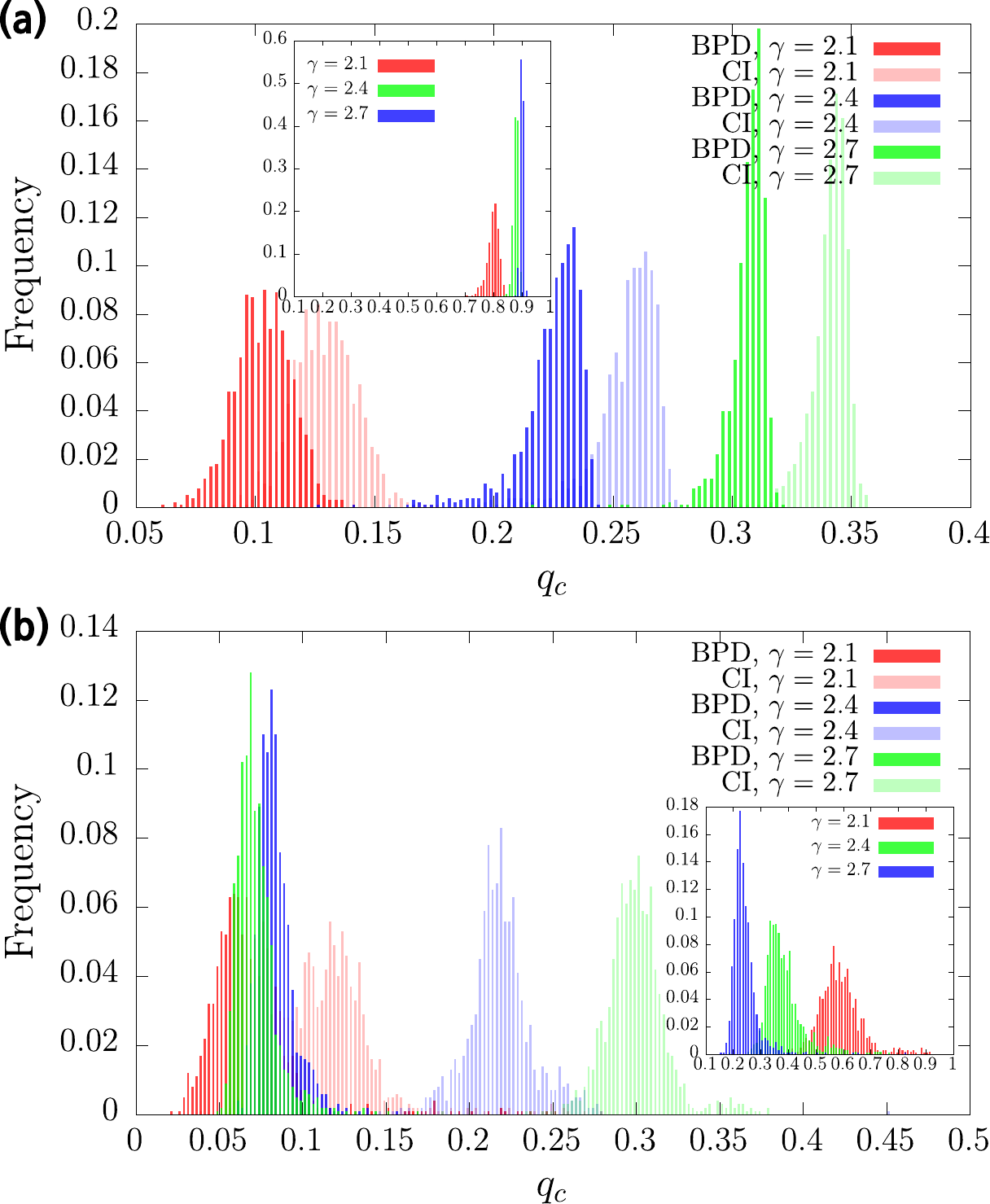}
\caption{Performance comparison of the BPD and CI algorithms on networks with different degree exponents $\gamma$. (a) the distribution of optimal percolation thresholds for $10^3$ realizations of the FNM with rewiring ratio $r=1$. (b) $10^3$ realizations of the FNM with $r=0$. The system size (number of nodes) of each realization ranges between $20000$ and $22000$, and the shortcut parameter is set to $s=3.0$.}
\label{fig:comparison-gamma}
\end{figure}

\subsection{Dependence on degree distribution}

The performance of dismantling algorithms also depends on the degree distribution of networks. When the shortcut rewiring ratio is fixed at $r=1$, the model networks are non-fractal for any choice of $2\leq\gamma\leq 3$. Thus, on the basis of the previous results, we expect that the distributions of $q_c$ for the BPD and CI algorithms are not greatly separated. Figure~\ref{fig:comparison-gamma}(a) confirms this expectation. The inset also shows that the performance gap between the BPD and CI algorithms is not very large for non-fractal networks with various degree exponents $\gamma$. For larger $\gamma$, the $q_c$ distributions for the BPD and CI algorithms look more similar. 
On the other hand, the performance gap between these algorithms varies with the degree exponent $\gamma$ at $r=0$. When $\gamma=2.1$, the two distributions overlap to some extent even at $r=0$. As $\gamma$ is increased, the separation between the distributions of $q_{c,\text{BPD}}$ and $q_{c,\text{CI}}$ increases. The reason is that a model network possesses more obvious fractality at $r=0$ for larger $\gamma$. As the fractality become more evident, so does the performance gap between the algorithms.

\begin{figure}[t]
\includegraphics[width=1.0\linewidth]{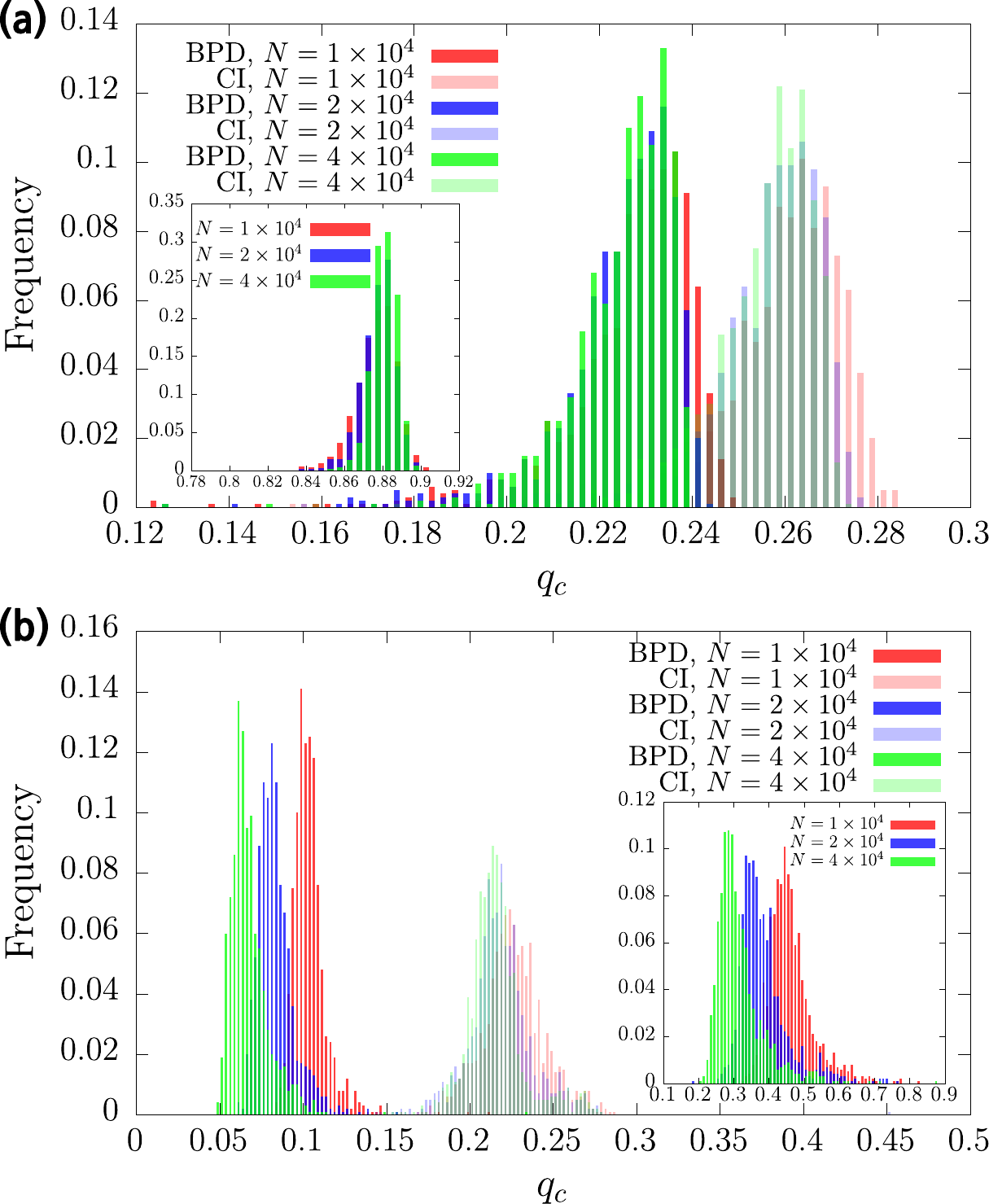}
\caption{Performance comparison of the BPD and CI algorithms on networks with different system sizes $N$. Because a critical branching tree is generated stochastically, the number of nodes in each tree can vary. It reveals that the number of nodes in each realization is between $1.0N$ and $1.1N$, where $N$ is given in the legend. The shortcut parameter is $s=3.0$, and the degree exponent is $\gamma=2.4$. (a) $r=1.0$ (non-fractal), (b) $r=0.0$ (fractal). Each distribution was obtained from $10^3$ realizations.}
\label{fig:comparison-size}
\end{figure}

\subsection{Dependence on the system size} 

Even though the system sizes $N$ are different, the distributions of $q_c$ for each algorithm have similar shapes for non-fractal networks ($r=1$, $\gamma\approx 2.4$) in Fig~\ref{fig:comparison-size}(a).  
For fractal networks ($r=0$, $\gamma\approx 2.4$), the BPD algorithm generates a smaller $q_c$ for larger $N$ in Fig.~\ref{fig:comparison-size}(b). The distribution of $q_c$ for the CI algorithm remains almost unchanged compared to that of the BPD algorithm.
As a result, the performance gap becomes larger when the two algorithms are applied on the fractal network of a larger system. Because we know that the FNM generates a network with more manifest fractality for larger $N$, this is consistent with the previous observation that the BPD algorithm is more effective than the CI algorithm, especially on fractal networks.

\begin{figure}[t]
\includegraphics[width=1.0\linewidth]{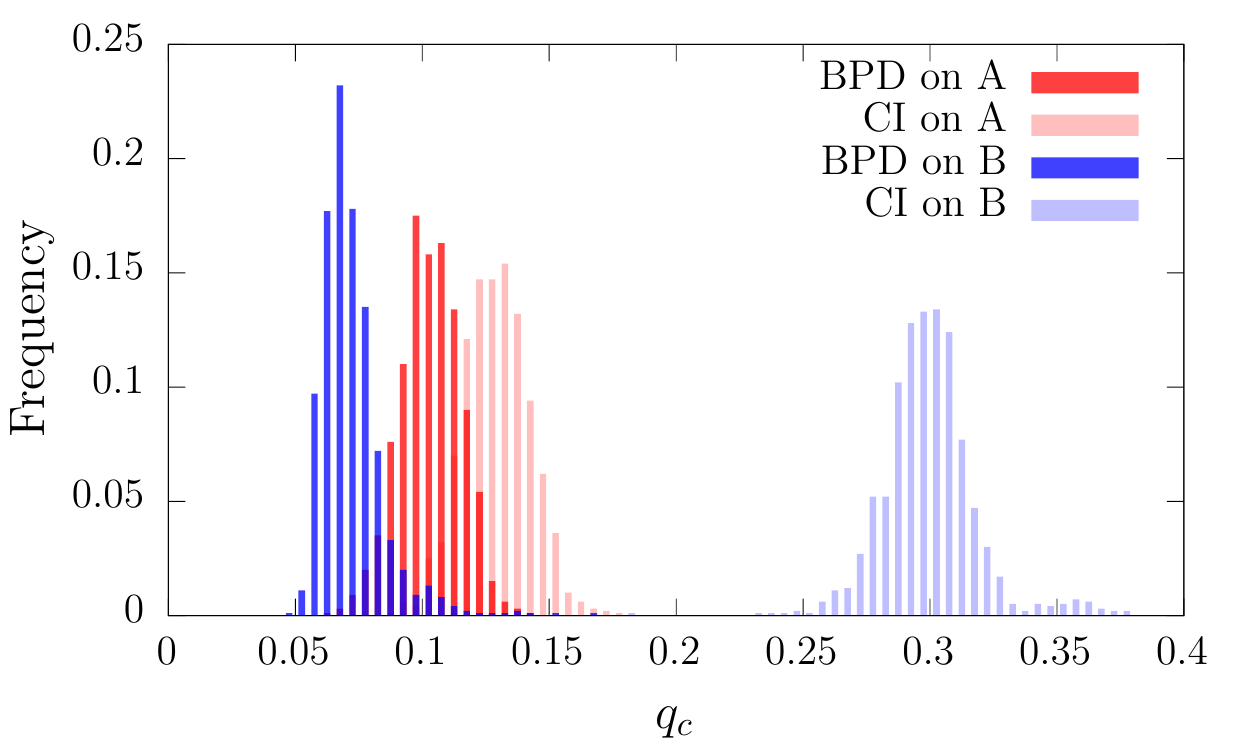}
\caption{Distribution of $q_c$ over sets of networks A and B found by BPD and CI algorithms. A is a set of $10^3$ realizations of the FNM with the degree exponent $\gamma\approx 2.1$ and the global shortcut parameters $s=3.0$ and $r=1.0$. B is a set of $10^3$ realizations of the FNM with $\gamma\approx 2.7$, $s=3.0$ and $r=0.0$.}
\label{fig:combination}
\end{figure}

\section{Reproduction of optimal percolation of real-world networks from model networks}

Using the above numerical results, we reproduce the change in the order of $q_c$ for two given types of networks when the two different dismantling algorithms are used. First, we prepare a pair of fractal and non-fractal networks with different degree exponents $\gamma$ but with the same other parameters as those in Fig.~\ref{fig:comparison-rewiring-ratio}. We obtain a wide (narrow) gap between the $q_c$ distributions for the BPD and CI algorithms on a fractal (non-fractal) network.  
If the degree exponents of the pair of networks were the same, we would reproduce the distributions that resemble those in Fig.~\ref{fig:comparison-rewiring-ratio}(b). If we can move the peaks of the $q_c$ distributions of the two algorithms on the non-fractal network ($r=1.0$) into the gap between the $q_c$ distributions on the fractal networks ($r=0.0$), the phenomena on the Web and the Internet are reproduced. Note that for non-fractal networks, the peaks of the $q_c$ distributions of the CI and BPD algorithms move to the left as the degree exponent decreases, as shown in Fig.~\ref{fig:comparison-gamma}(a). Using this observation, we set a smaller degree exponent for the non-fractal network than for the fractal network. Then, we can expect that the $q_c$ distributions of the two algorithms on non-fractal networks are located between the peaks of the $q_c$ distributions of the BPD and CI algorithms on fractal networks (Fig.~\ref{fig:combination}).
Thus, if we use the BPD algorithm on a fractal network with larger $\gamma$ (one realization of Fig.~\ref{fig:combination}(b)) and a non-fractal network of smaller $\gamma$ (Fig.~\ref{fig:combination}(a)), $q_c$ is smaller for the fractal network than for the other network with high probability. Further, if we apply the CI algorithm on the same pair of networks, the algorithm finds that the non-fractal network with smaller $\gamma$ has a smaller $q_c$ than the fractal network.
Thus, if we have a fractal network with large $\gamma$ and a non-fractal network with small $\gamma$, the BPD and CI algorithms will disagree about the position of $q_c$. This behavior is similar to that obtained from the two real-world networks, the World Wide Web and Internet.

\section{Conclusion and discussion}

We studied the behavior of two well-known heuristic algorithms for the optimal percolation problem, the CI and BPD algorithm, on networks with many loops. Model networks were generated from the FNM, and the loop characteristics were controlled by varying the shortcut rewiring parameter $r$. For $r=0$, most shortcuts were drawn locally, and the network contained many local loops. For $r=1$, where the resultant network became clearly non-fractal regardless of the degree exponent, every shortcut was rewired randomly. Both algorithms were developed assuming that the target network is a random sparse network that is locally treelike; thus, both algorithms are expected to work well on non-fractal networks, but the performance on fractal networks is not guaranteed.

In every graph instance, the BPD algorithm gave better results. The performance of the CI algorithm was comparable to that of the BPD algorithm on non-fractal networks, but the performance gap between the BPD and CI algorithms became very large on fractal networks. Fractal networks maintain hierarchical structure originating in the critical branching tree; thus, we can expect that they should have a much smaller $q_c$, as the distribution of $q_{c,BPD}$ is small. However, it seems that the CI algorithm fails to capture the fragile structure of fractal networks (Fig.~\ref{fig:comparison-rewiring-ratio}(b)).

This result can lead to interesting behavior such as that we found in a pair of real-world networks, subgraphs of the Web and the Internet (Fig.~\ref{fig:comp-real}). For scale-free networks with similar number of links, a network with a smaller degree exponent $\gamma$ is generally harder to destroy (Fig.~\ref{fig:comparison-gamma}). Further, rewiring of shortcuts connects distinct local communities much more closely, making the network more robust (Fig.~\ref{fig:comparison-rewiring-ratio}). If we prepare a scale-free network A with a high degree exponent and locally drawn shortcuts (fractal), and a network B with a low degree exponent and randomly drawn shortcuts (non-fractal), the effects of the degree exponent $\gamma$ and rewiring parameter $r$ compete with each other. If the the number of links is sufficiently large, the effect of rewiring can surpass the effect of the degree exponent, making fractal network A more fragile than non-fractal network B even though B has a smaller degree exponent. However, the CI algorithm finds that B is more fragile, because the CI algorithm performs poorly on fractal networks. As a result, the two algorithms disagree about which network is more robust (Fig.~\ref{fig:combination}).

The gap between the results of the CI and BPD algorithms can represent how much a network differs from a sparse random network that is locally treelike. Braunstein et al. also reported a similar observation in the appendix of \cite{Braunstein2016}, showing that the CI algorithm (without an additional revival process) performs relatively poorly on Watts--Strogatz (WS) networks with a small rewiring probability, and its performance becomes comparable to that of other algorithms on WS networks with a large rewiring probability, which are sparse random networks. From this viewpoint, it makes sense that the CI algorithm does not work well on fractal networks. Fractal networks can be seen as a class of networks that are very different from locally treelike networks, as fractal networks maintain the hierarchical structure of a branching tree and contain many local loops. 

Changes in the gap between the results of the BPD and CI algorithms due to variations in the parameters can be explained in this sense. Given the same node number $N$, shortcut parameter $s$, and rewiring ratio $r$, a network generated by the FNM is far from a fractal network if the degree exponent $\gamma$ is small (Fig.~\ref{fig:FS-model}). This results in a smaller gap between the results of the CI and BPD algorithms for smaller $\gamma$ (Fig.~\ref{fig:comparison-gamma}). In addition, for $r=0$, the network is less locally treelike as the shortcut parameter $s$ increases, causing a larger gap between the two algorithms. On the other hand, a larger shortcut parameter $s$ makes the graph more random when we allow rewiring. Thus, the results of the CI algorithm become more comparable to those of the BPD algorithm on networks with larger $s$, which are globally rewired ($r=0.5, 1.0$). Most importantly, controlling the rewiring ratio $r$ affects the fractality directly, so the gap between the CI and BPD algorithms also varies significantly.

Fractal scale-free networks constitute only a small portion of the entire family of random loopy networks, but they can be generated and understood systematically and are also easily found in the real world \cite{Goh2006}. Studying the optimal percolation problem on fractal networks will be helpful for designing new algorithms that are effective on networks with many local loops.

\begin{acknowledgments}
This work was supported by the National Research Foundation of Korea by Grant No. NRF-2014R1A3A2069005.
\end{acknowledgments} 

\appendix

\section{Collective influence algorithm}
Although the CI algorithm~\cite{Makse2015} starts theoretically from sophisticated considerations of local stability analysis of message-passing equations, it provides a simple centrality measure as a criterion for selection of nodes to be deleted. Assuming that the network is locally treelike, the solution with a vanishing giant connected component depends on the largest eigenvalue of the modified non-backtracking matrix. The solution is stable only when the eigenvalue is less than unity, whereas the eigenvalue drops abruptly from one to zero when no loop remains. By using a perturbative method, the problem is reduced to minimizing the cost function, which is defined as the sum of CI$_\ell(i)$ over all nodes, where
\begin{align*}
\text{CI}_\ell(i)=(k_i-1)\sum_{j \in \partial {\rm Ball}(i,\ell)}(k_j-1).
\end{align*}
The CI algorithm repeatedly removes the node with the largest CI value until the largest connected component vanishes. The CI value of every node is re-evaluated after each removal. The algorithm outperforms intuitive decimations based on traditional centrality measures such as the degree or eigenvalue centrality, because it can take into account the importance of \textit{weak nodes} with small degrees. Although the algorithm becomes exact as $\ell \to \infty$ for an infinite treelike network, a small $\ell$ still yields good estimation for finite networks. Moreover, deleting a fixed fraction of nodes with the largest CI values at once does not affect the performance in typical cases, allowing the algorithm to work in a time complexity of $O(N\log N)$. In this study, the CI algorithm uses CI$_{\ell=2}$ as its criterion because this value is more effective than larger or smaller $\ell$ for the prototypical system size of the model networks used here. When we take an excessively large $\ell$ value, the computation time becomes long and the performance is degraded, because the algorithm performs only random deletion when $\ell$ is equal to or larger than the network diameter. A $0.1\%$ fraction of the nodes of the original network were deleted at each step until the size $G$ of the giant component reached $1\%$ of the number of nodes of the original network.

\section{Belief-propagation-guided decimation algorithm}
The BPD algorithm \cite{Zhou2016} uses the minimum feedback vertex set (mFVS) problem as an approach to the optimal percolation problem. The mFVS problem is to find a minimal set of nodes whose removal eliminates every loop. If we draw a subgraph on the original graph by retaining only nodes that have one parent, the subgraph consists only of simple loops and trees~\cite{Zhou2013}. This rule is local, so it can be expressed by BP equations involving the variables of neighboring nodes. The algorithm evaluates the marginal empty probability $q_0^i$ of each node at each moment from the probabilities in the cavity graphs, which are calculated by iteration of the BP equations. Definitions of $q_0^i$ and the BP equations can be found in \cite{Zhou2016}. The node with the highest $q_0^i$ is removed because it is strongly recommended to remove that node in order to draw a subgraph without loops. Although the BP equations are not guaranteed to converge to a fixed point on general graphs, the equations are iterated a fixed number of times in this algorithm. In practice, multiple nodes with the highest $q_0^i$ are deleted together in one step. After the evaluation of $q_0^i$ and the node removal, another cycle of iteration and node removal is repeated until no loop remains. 

The resultant tree components are broken into pieces by removing additional nodes until no remaining connected component is larger than expected. Because our purpose is to decompose the giant component, small components with loops can be allowed. Thus, among the deleted nodes, some nodes are revived unless a large component emerges upon their revival. The BPD algorithm reportedly outperforms the CI algorithm on various types of models and real-world networks~\cite{Zhou2016}. The BPD algorithm is based on the spin glass model, where each possible microscopic state can be realized by a probability weighted by the number of remaining nodes multiplied by the reweighting parameter $X$. In this study, the reweighting parameter $X$ is set to $12.0$. When it is sufficiently large, it does not affect the performance of the algorithm significantly, even though the shape of the curve for each trial can vary slightly. At each step, the BP equations are iterated, and $1\%$ of the remaining nodes are deleted. Then, the remaining tree components are broken into pieces by deleting additional nodes until the size $G$ of the giant component reaches $1\%$ of the number of nodes in the original network.

\end{document}